\documentclass[a4paper,aps,pre,floatfix,twocolumn,nofootinbib,showpacs,superscriptaddress]{revtex4-1}

\usepackage[pdftex]{graphicx}
\usepackage{amssymb,latexsym,amsmath}
\usepackage[pdftex]{color}
\usepackage{hyperref}
\usepackage{float}
\usepackage{bm}
\usepackage{bbm}
\usepackage{xspace}
\usepackage{multirow}

\newcommand{\av}[1]{\langle #1 \rangle}

\newcommand{\eps}{\varepsilon}

\begin{document}

\title{Scale-free networks emerging from multifractal time series}

\author{Marcello A. Budroni}

\affiliation{Nonlinear Physical Chemistry Unit, Facult\'e des Sciences,
  Universit\'e libre de Bruxelles (ULB), CP 231 - Campus Plaine, 1050
  Brussels, Belgium.}

\author{Andrea Baronchelli}

\affiliation{Department of Mathematics - City, University of London -
  Northampton Square, London EC1V 0HB, UK}

\author{Romualdo Pastor-Satorras}

\affiliation{Departament de F\'\i sica, Universitat Polit\`ecnica de
  Catalunya, Campus Nord B4, 08034 Barcelona, Spain}

\date{\today}

\begin{abstract}
  Methods connecting dynamical systems and graph theory have attracted
  increasing interest in the past few years, with applications ranging
  from a detailed comparison of different kinds of dynamics to the
  characterisation of empirical data. Here we investigate the effects of
  the (multi)fractal properties of a time signal, common in sequences
  arising from chaotic or strange attractors, on the topology of a
  suitably projected network. Relying on the box counting formalism, we
  map boxes into the nodes of a network and establish analytic
  expressions connecting the natural measure of a box with its
  degree in the graph representation. We single out the conditions
  yielding to the emergence of a scale-free topology, and validate our
  findings with extensive numerical simulations.

\end{abstract}


\maketitle

\section{Introduction}
\label{sec:introduction}

Network science \cite{Newman2010} has emerged in the last decades as a
transverse interpretative framework for understanding the structure and
function of a wide range of complex systems, as well as the dynamic
phenomena that takes place on them, ranging from financial crises and
traffic congestion to epidemic and social influence spreading
\cite{barratbook,dorogovtsev07:_critic_phenom,Pastor-Satorras:2014aa}.

In the realm of dynamical systems \cite{Guckenheimer2002}, network
techniques have been applied to the analysis of nonlinear
time series, with a particular focus on characterizing chaotic dynamics
\cite{ottchaos}. The main idea of this methodology is to project the
information of a time series into the topology of a network. The key
element of this approach resides in the identification of nodes and
links in the network from the time series information. Several
alternatives have been proposed in this context
\cite{doi:10.1142/S0218127411029021}. Thus, Zhang and
Small~\cite{Li2012} consider cycles of a pseudo-periodic time series as
the nodes of a network, which are connected by links depending on the
similarity between cycles.  Lacasa \textit{at al.}
\cite{Lacasa01042008,ISI:000282271800002}, on the other hand, build on
the concept of \textit{visibility graphs}, where nodes correspond to
series data points and two nodes are connected if a straight line can be
established between them without intersection with any intermediate data
height.

Another general approach is based on encoding topological information
from the reconstructed phase space of a time series into a
\textit{proximity network}. In these networks, nodes represent segments
of time series or vectors in the related reconstructed phase space, and
links depend on a specific criterion determining adjacency in phase
space. Cycle networks \cite{PhysRevLett.96.238701}, correlation networks
\cite{Yang20081381,PhysRevE.79.066303}, and recurrence networks
\cite{Marwan20094246,gao2009,1367-2630-12-3-033025,xiang2012} are
typical examples of proximity networks. Finally, the related class of
\textit{transition networks}
\cite{doi:10.1142/S0218127405014167,1742-5468-2009-07-P07046,Campanharo2011,sun2014}
encompasses different models for mapping time series into networks, in
which the values on the time signal are mapped into a finite number of
states (regions in the phase space of the series), representing the
nodes, which are connected if the signal transits from one state to
another.  This transformation procedure preserves the temporal
information of the dynamics in the network, is stable to noise affecting
real time series and can account for the fundamental structure of the
related attractors \cite{sun2014}.

Using this kind of network projections, several aspects of dynamical
systems have been cast in terms of topological network properties. Here
we will focus in particular on the effects that the \textit{fractal} and
\textit{multifractal} properties of a time signal have on the topology
of representative classes of projected network. Many time signal,
particularly those arising from a chaotic or strange attractor
\cite{ottchaos}, have a fractal structure, characterized by a
statistically self-similar pattern in phase space. Moreover, they can
also show multifractal properties, described by a strongly heterogeneous
probability of visiting different neighborhoods in the phase space
\cite{PhysRevA.33.1141,Falconer:2003fk,feder2013fractals}.  Using as a
simple example a transition network representation
\cite{1742-5468-2009-07-P07046,Budroni20103883}, framed within a box
counting formalism \cite{Falconer:2003fk} in which each box corresponds
to a vertex, we find analytic expressions tying the visitation
probability of a box with its associated degree. From these relations we
obtain the conditions under which a projected network grows with a
scale-free topology \cite{Barabasi:1999}, characterized by a degree
distribution of the form $P(k) \sim k^{-\gamma}$. Our results highlight
the correspondence between an attractor's structure and the topology of
the associated projected network, and relate the possible origin of a
heterogeneous scale-free topology with the heterogeneous and
hierarchical visitation probability characterizing multifractal
attractors.

The present paper is organized as follows: In
Sec.~\ref{sec:mult-time-sign} we briefly summarize the multifractal
formalism for chaotic time series and attractors. In
Sec.~\ref{sec:nework-mapping} we present a transition network mapping
for general time series, based on the box-counting algorithm
\cite{1742-5468-2009-07-P07046}. In Sec.~\ref{sec:topol-prop-proj} we
relate the topological properties of the associated networks with the
multifractal properties of the original time series. This relation is
directly mediated by the natural measure of the series, defined as the
probability that the time sequence visits a given box in a partition of
the substrate of the series. Numerical checks of our theoretical
predictions are detailed in Sec.~\ref{sec:numer-exper}. Finally, we
present our conclusions in Sec.~\ref{sec:discussion}.

\section{Multifractal time signals}
\label{sec:mult-time-sign}

For time signals arising from strange (chaotic) attractors, it is common
that different regions of the phase space are differently visited, and
chaotic orbits spend most of their time in a small region of the support
underneath the chaotic attractor itself.  This heterogeneity is at the
basis of the so-called multifractal structure of the strange attractor,
which can be mathematically captured by the formalism presented below
\cite{PhysRevA.33.1141,feder2013fractals}.

Let us consider a real time signal, or a $d$-dimensional chaotic
attractor, given by the normalized sequence of points
$\mathcal{G} = \left\{ \vec{x}_t : \vec{x}_t \in [0,1]^d, t= 1, 2,
  \ldots n \right \}$, where the index $t$ represents either time or the
order in the sequence of points that generate the attractor (e.g. the
index of the iteration in an iterated map), while $n \gg 1$ is the
number of points in the signal. We consider a partition of the set
$[0,1]^d$ in $M$ boxes of length $\varepsilon = M^{-1/d}$. Box are
labeled by the indexes $i$, with $1 \leq i \leq M$.  Let us associate to
each point $\vec{x}_t = \{ x_t^{(1)}, x_t^{(2)}, \ldots, x_t^{(d)}\}$ in
the sequence an integer index
\begin{equation}
  \label{eq:6}
  i_{t} = 1+ \sum_{r=1}^{d} \left \lfloor x_t^{(r)} L \right \rfloor \;
  L^{r-1}
\end{equation}
in the range $1 \leq i_t \leq M$, where $L = M^{1/d}$, and
$\lfloor z \rfloor$ is the floor function.  In this sense, the signal or
attractor can be interpreted as visiting the $i_t$-th box in the
partition at time $t$.  In heterogeneous fractals, not all the boxes will be equally visited. In general, during the $n$ steps of the signal, the $i$-th box will be
visited a number of times $n_i$, and the total number of boxes visited
at least once will be $N(\eps)$. This quantity coincides with the number
of boxes of length $\eps$ needed to cover the fractal set, and thus we
can define the box or capacity dimension of the attractor
\cite{Falconer:2003fk}, $D_0$, by the relation
\begin{equation}
  \label{eq:8}
  N(\eps) \sim \eps^{-D_0}.
\end{equation}

Let us define the probability $p_i(\eps) = \lim_{n \to \infty} n_i /n$,
termed the \textit{natural measure}, as the probability that the chaotic
map visits the $i$-th box of the $N(\eps)$ available during an
infinitely long orbit.  For an homogeneous structure in $d$ dimensions,
$p_i(\eps) \sim \eps^d$, while in the case of a uniform fractal of
dimension $D_0 \leq d$, $p_i(\eps) \sim \eps^{D_0}$.  In more complex
situations, however, the attractor exhibits a non-uniform fractal
distribution, and we assume a general form for the natural measure
$p_i(\eps) \sim \eps^{\alpha_i}$, where the exponent $\alpha_i$, taking
values in the interval $[\alpha_\mathrm{min}, \alpha_\mathrm{max}]$,
measures the strength of the local singularity of the measure at box
$i$. In general, there will be many boxes with the same value of
$\alpha$, such that their number scales as
$N_\alpha(\varepsilon) \sim \varepsilon^{-f(\alpha)}$.  The function
$f(\alpha)$, called the multifractal spectrum of the measure, defines
the fractal dimension of the set of boxes with the given value $\alpha$,
and is in general a convex function with a single maximum. An
equivalent, and numerically simpler, description can be obtained from
the generalized dimensions $D_q$, defined as \cite{PhysRevA.33.1141}
\begin{equation}
  \label{eq:1}
  D_q = \frac{1}{q-1} \lim_{\eps\to0}\frac{ \log \sum_i
    p_i(\eps)^q}{\log \eps},
\end{equation}
which, for $q \geq 0$, fulfill
$D_0 \leq D_q \leq D_\infty \equiv \alpha_\mathrm{min}$.  For a uniform
measure $D_q=\mathrm{const.}=D_0$. For multifractal measures, $D_q$ is a
decreasing function of $q$ which is related with the multifractal
spectrum ($\alpha, f(\alpha)$) by means of a Legendre transformation
\cite{PhysRevA.33.1141}, defining a parametric exponent $\alpha(q)$ that
fulfills the equations
\begin{eqnarray}
  \label{eq:2}
  \alpha(q) &=& \frac{d}{d q} (q-1)D_q \\
  f(\alpha(q)) &=& q \alpha(q) + (1-q) D_q
\end{eqnarray}
Numerically, the generalized dimensions can be estimated from
Eq.~(\ref{eq:1}), by noticing that, for finite $\eps$,
\begin{equation}
  \label{eq:9}
  \sum_i p_i(\eps) \sim \eps^{(q-1)D_q},
\end{equation}
allowing $D_q$ to be determined from a linear regression of
$\sum_i p_i(\eps)$ for decreasing values of $\eps$ (increasing $M$) in a
log-log plot.

\section{Nework mapping}
\label{sec:nework-mapping}

In order to construct a transition network representation, we follow the
approach in Refs.~\cite{1742-5468-2009-07-P07046,Budroni20103883}, and
associate a (virtual) vertex to each box $1 \leq i \leq M$ in the
partition of the chaotic attractor in phase space. Actual vertices in
the network are given by the set of boxes that have been visited at
least once by the signal, with a size $N(\eps)$. Edges in the network
are established by associating an undirected connection between vertices
$i$ and $i'$ whenever the signal jumps between boxes $i$ and
$i'$ in two consecutive time steps, i.e. $i \equiv i_t$ and
$i' \equiv i_{t+1}$.

The resulting projected networks, which are characterized by the
coarse-grained scale $\eps$, are connected by construction, and preserve
temporal information of the generator of the signal.  A completely
random, stochastic signal will lead to a fully connected network; on the
other hand, for a limit cycle or periodic attractor, the projected
network will be ring, with a number of nodes equal to the period of the
cycle. The former case corresponds to $D_0 = d$, and the latter to
$D_0 = 1$.

\section{Relating topology with multifractality}
\label{sec:topol-prop-proj}

In the case of a multifractal time series, the topology of the
associated transition network described above can be related to the
generalized dimensions $D_q$. We will consider in particular the degree
distribution $P_\eps(k)$ of a projected network with a coarse-graining
level $\eps$, defined as the probability that a randomly chosen node has
degree $k$, i.e., it is connected to $k$ other nodes.  To make explicit
this relation, we observe that every node (box) $i$, will be
characterized by a number of visits $n_i$ and a degree $k_i$. In the
limit $n\to\infty$, we assume that the relative number of visits,
i.e. the natural measure $p_i(\eps)$, and the degree of a node are
stable quantities. The corresponding degree distribution will thus
depend only on the phase space discretization $\eps$. On average, the
two quantities $p_i(\eps)$ and $k_i$ will be related, since obviously
the more times a box is visited, the larger is expected to be the degree
of the associated node. We can thus assume a functional relation between
these two averaged quantities, valid for sufficiently large $k_i$ and
$p_i$, of the form
\begin{equation}
  \label{eq:13}
  p_i(\eps) \simeq   G_\eps(k_i),
\end{equation}
where $G_\eps(z)$ is an increasing function of $z$.  The function
$G_\eps(k_i)$ depends in general on $\eps$. Indeed, from the
normalization of the natural measure, Eq.~(\ref{eq:13}) implies that
\begin{equation}
  \label{eq:10}
  \sum_i G_\eps(k_i) \equiv N(\eps)\sum_k P_\eps(k) G_\eps(k) =
  N(\eps)\av{G_\eps(k)} =1,
  \nonumber
\end{equation}
where we have defined $\av{F(k)} = \sum_k P_\eps(k) F(k)$. From here, we
have
\begin{equation}
  \label{eq:11}
  \av{G_\eps(k)} =  N(\eps)^{-1} \sim \eps^{D_0}.
\end{equation}
We make the assumption, to be validated numerically later on, that the
$\eps$ dependence of the function $G_\eps(k)$ resides exclusively in a
multiplicative prefactor, i.e.  $G_\eps(k) = a(\eps) g(k)$. From
Eq.~(\ref{eq:11}), we have that
$ \av{G_\eps(k)} = a(\eps) \av{g(k)} \sim \eps^{D_0}$. We make the
additional assumption that the average of $g(k)$, $\av{g(k)}$, is a
constant, independent of $\eps$. From here, we obtain the relation
\begin{equation}
  \label{eq:12}
  p_i(\eps) \simeq  \eps^{D_0} \, g(k_i).
\end{equation}
This relation implies that we can express the multifractal properties of
the attractor in terms of topological properties of the network. In
fact, from Eq.~(\ref{eq:12}), we have
\begin{eqnarray}
  \label{eq:15}
  \sum_i p_i(\eps)^q &\simeq& \sum_i \left[
                              \eps^{D_0} g(k)  \right]^q \simeq
                              \eps^{qD_0} N(\eps) \sum_k P_\eps(k)
                              g(k)^q \nonumber\\
                     & \simeq &  \eps^{(q-1)D_0}  \av{g(k)^q},
\end{eqnarray}
where we have used $N(\eps) \sim \eps^{-D_0}$.  From Eq.~(\ref{eq:9})
we have also $\sum_ip_i(\eps)^q \sim \eps^{(q-1)D_q}$. Combining this
with Eq.~(\ref{eq:15}), we obtain the relation linking network and
multifractal properties, namely
\begin{equation}
  \label{eq:17}
  \av{g(k)^q} \sim \eps^{-(q-1)(D_0-D_q)}.
\end{equation}
For a homogeneous fractal set, $D_q = D_0$ $\forall q$, and
$ \av{g(k)^q} =\mathrm{const.}$ On the other hand, for a multifractal
strange attractor, since $D_q < D_0$ for $q>0$, we have that the moments
$\av{g(k)^q}$, with $q > 1$, diverge as the number of boxes in the
partition increases, i.e. as the network size grows. This observation
allows to extract conclusions on the functional form of the degree
distribution, which will depend on the particular growth law $g(k)$. We
will consider analytically tractable forms in the following Section.  To
avoid complications in the development, we will further assume in our
analysis that the degree distribution of the projected network is
stable, meaning that the effect of $\eps$ consists essentially in
imposing an upper degree cut-off $k_c$ to a functional form $P(k)$
independent of $\eps$.

\subsection{Exponential growth}
\label{sec:exponential-growth}

Let us consider first the case of an exponential (faster than algebraic)
growth of the number of visits in a box with the associated node degree,
i.e.
\begin{equation}
  \label{eq:18}
  g(k) \sim e^{\beta k},
\end{equation}
where $\beta >0$. The fact that $\av{e^{\beta k}}$ is constant and
$\av{e^{\beta q k}}$ diverges for $q>1$ is compatible with a degree
distribution that shows, at large values of $k$, the asymptotic behavior
$P(k) \sim e^{-\alpha k}$, with $\alpha>\beta$, that is, an
exponentially bounded degree distribution. Assuming a stable degree
distribution, for a non-zero $\eps$ the divergence of the exponential
moments $\av{e^{\beta q k}}$ will be reflected in a dependence on the
network size $N(\eps)$, modulated by the largest degree in the network,
or degree cut-off $k_c(\eps)$ \cite{mariancutofss}. To estimate this
value, we observe that, from Eq.~(\ref{eq:12}),
$k_i \sim \ln \left[p_i(\eps)/\eps^{D_0}\right]^{1/\beta} \sim \ln
\left[ \eps^{ -(D_0 - \alpha_i) / \beta} \right]$. The largest value of
$k_i$ will correspond to the minimum of $\alpha_i$,
$\alpha_\mathrm{min} = D_\infty$. Therefore, we have
\begin{equation}
  \label{eq:3}
  k_c(\eps) 
  \sim \ln \left[ \eps^{ -(D_0 - D_\infty) / \beta} \right] .
\end{equation}
This expression allows to relate the network parameters $\alpha$ and
$\beta$ with the multifractal exponents $D_0$ and $D_\infty$ by noticing
that, in a network of finite size $N(\eps)$, the maximum degree is given
by the condition $\sum_{k=k_c(\eps)}^\infty P(k) = 1/N(\eps)$
\cite{mariancutofss}.  With an exponential degree distribution
$P(k) = \alpha e^{-\alpha k}$ we thus have, in the continuous degree
approximation,
$\int_{k_c(\eps)}^\infty \alpha e^{-\alpha k} \; dk = e^{-\alpha
  k_c(\eps)} = 1/N(\eps)$, from where we obtain
\begin{equation}
  \label{eq:5}
   k_c(\eps)  \sim \ln \left[N(\eps)^{1/\alpha}\right]  \sim \ln
   \left[\eps^{-D_0/\alpha} \right].
\end{equation}
Combining Eqs.~(\ref{eq:3}) and~(\ref{eq:5}), we obtain the relation
\begin{equation}
  \label{eq:20}
  \frac{\alpha}{\beta} = \frac{D_0}{D_0 - D_\infty}.
\end{equation}

The properties of the network can also be used to extract information on
the full set of generalized dimension by building on relation
Eq.~(\ref{eq:17}). Indeed we can write the diverging moments in a finite
network as
\begin{eqnarray}
  \label{eq:4}
  \av{e^{\beta q k}} &\sim& \int^{k_c(\eps)} e^{(q \beta - \alpha)k} \;
  dk \sim \frac{ e^{(q \beta - \alpha)k_c(\eps)}}{q \beta - \alpha}
                            \nonumber \\
\end{eqnarray}
which, taking into account Eq.~(\ref{eq:3}), yields

\begin{eqnarray}
  \label{eq:4bis}
  \av{e^{\beta q k}} & \sim&\eps^{-(D_0 - D_\infty)(q - \alpha/\beta)} \, .
\end{eqnarray}

Combining Eq.~(\ref{eq:4bis}) and Eq.~(\ref{eq:17}) leads to the asymptotic
expression, valid for large $q$
\begin{equation}
  \label{eq:21}
  D_q \sim  D_\infty \frac{q }{q-1},
\end{equation}
that recovers the result known for deterministic multifractal measures
\cite{feder2013fractals}.

\subsection{Algebraic growth}
\label{sec:algebraic-growth}

In the case of an algebraic growth of the number of visits with the
associated degree, we have
\begin{equation}
  \label{eq:19}
  g(k) \sim k^\delta,
\end{equation}
with $\delta >0$. From Eq.~(\ref{eq:17}), we have that $\av{k^\delta}$
is finite, and $\av{k^{q \delta}}$, with $q>1$, diverge in the limit of
infinite network size.  The fact that all higher degree moments diverge
indicate that the degree distribution $P(k)$ of the network has long
tails, which in the simplest case are compatible with a power law
distribution of the form $P(k) \sim k^{-\gamma}$, where we impose
$\gamma > \delta+1$ to ensure a finite value of $\av{k^\delta}$.

Performing again a finite-size analysis, from Eq.~(\ref{eq:12}) we obtain
a maximum degree
\begin{equation}
  \label{eq:22}
  k_c(\eps) \sim \eps^{-(D_0-D_\infty)/\delta} \sim N(\eps)^{(1-D_\infty/D_0)/\delta}.
\end{equation}
On the other hand, the network relation
$\sum_{k=k_c(\eps)}^\infty P(k) = 1/N(\eps)$ leads now, with a power-law
degree distribution $P(k) \sim k^{-\gamma}$ in the continuous degree
approximation, to \cite{mariancutofss}
\begin{equation}
  \label{eq:23}
  k_c(\eps) \sim N(\eps)^{1/(\gamma-1)} \sim \eps^{-D_0/(\gamma-1)}.
\end{equation}
Combining Eqs.~(\ref{eq:22}) and~(\ref{eq:23}), we obtain the relation
between network properties and multifractal exponents
\begin{equation}
  \label{eq:24}
  \frac{\gamma-1}{\delta} = \frac{D_0}{D_0 - D_\infty}.
\end{equation}
In the case of an algebraic $g(k)$ function, Eqs.~(\ref{eq:22})
and~(\ref{eq:23}) can be used to directly estimate $\delta$ and $\gamma$. This
approach is more difficult in the case of an exponential $g(k)$, due to
the much smaller range of variation of the logarithm of $\eps$, see
Eqs.~(\ref{eq:3}) and~(\ref{eq:5}).

Finally, writing
\begin{eqnarray}
  \label{eq:25}
  \av{k^{\delta q}} &\sim& \int^{k_c(\eps)} k^{\delta q - \gamma} \;dk \sim
  k_c(\eps)^{\delta q - \gamma+1} \nonumber \\
                   &\sim& \eps^{-D_0[\delta q /(\gamma-1) -1]},
\end{eqnarray}
and comparing with Eq.~(\ref{eq:17}), we obtain again the simple
asymptotic expression for the generalized dimensions given by
Eq.~(\ref{eq:21}).

\section{Numerical experiments}
\label{sec:numer-exper}

In order to check the validity of the predictions made in
Section~\ref{sec:topol-prop-proj}, we have considered different
multifractal time signals, generated by means of iterative maps. In particular we have studied three paradigmatic examples of 1- and 2-dimensional chaotic attractors, namely
the logistic, the Duffing and the Henon map.  The well-known logistic recurrence \cite{may76,ottchaos} in dimension $d=1$,
\begin{eqnarray}
  x_{t+1} &=&  \mu \, x_t (1-x_t)  \, ,
\end{eqnarray}
maps the interval $x \in \left[0, 1\right]$ into itself when the control
parameter $\mu$ ranges between 0 and 4. This systems undergoes a period-doubling bifurcation transition to chaos, which sets-in at $\mu = 3.569456 ...$ . Multifractal chaotic regimes
interspersed with periodic windows then occur in the parameter interval
$\mu \in [3.57, 4)$.  Here we fix $\mu=3.7$.  The two-dimensional
Duffing map
\begin{eqnarray}
  x^{(1)}_{t+1} &=&  x^{(2)}_t \, , \\
  x^{(2)}_{t+1} &=&  - b\,x^{(1)}_{t} + a \,x^{(2)}_{t} - \,\left[x^{(2)}_{t}\right]^3\,,
\end{eqnarray}
is a discrete representation of the Duffing oscillator, describing a forced oscillator coupled to a dissipative restoring force \cite{ottchaos}. This map typically produces chaotic behaviours with the critical parameters $a=2.75$ and $b=0.2$, and generates values of $x^{(1)}$ and
$x^{(2)}$ in the range $[-1.71, 1.71]$. Following our network projection algorithm, each variable is thus shifted
and normalized into the interval $\left[0,1\right]^2$.  Finally, the
Henon map \cite{henon,ottchaos} in $d=2$ is defined by the recurrence
\begin{eqnarray}
  x^{(1)}_{t+1} &=&  1 - a\left[x^{(1)}_t\right]^2 + x^{(2)}_t \, , \\
  x^{(2)}_{t+1} &=&  b\,x^{(2)}_{t} \,,
\end{eqnarray}
with the parameters $a$ and $b$ fixed to $1.4$ and $0.3$,
respectively. For these values, starting from an initial point ($x^{(1)}_0,x^{(2)}_0$) the dynamics can either asymptotize to a fractal attractor relying on the subset $x^{(1)} \in [-1.3,1.3]$ and $x^{(2)} \in [-0.4,0.4]$ or diverge to infinity. For other values of $a$ and $b$ the map may be still chaotic, intermittent, or converge to a periodic orbit. 
Here we use the classical parameter setting and, again, we transform the values of each variable into the interval $\left[0,1\right]^2$. A
one-dimensional projection of the Henon map is also considered, obtained
by taking into account only one normalized variable of the map (both $x^{(1)}$ and $x^{(2)}$ give analogous results).

\begin{table}[b]
  \centering
  \begin{tabular}{|c|c|cc|cc|cc|}
    \hline \hline
    Attractor & \xspace$d$ \xspace  & $D_0$    & $D_\infty$ & $\beta$ & $\delta$ & $\alpha$ & $\gamma$\\ \hline
    \hline
    Logistic  &$1$  & 0.998(2) & 0.49(1)
                                            & ---      &  1.04(2)& ---
                                                                            &  3.14(1) \\ \hline
    Duffing   &$2$  & 1.306(2) & 0.75(1)    & 0.28(8)  &  ---    & 0.42(6)  &   ---     \\ \hline
    \multirow{2}{*}{Henon}
              &$1$  & 1.000(1) & 0.73(2)    &  ---     &  1.10(5)&     ---  &  4.48(1) \\
              &$2$  & 1.24(1)  & 0.82(2)    & 0.15(2)  & ---     & 0.57(5)     &   ---
                                                                                       \\\hline
    \hline
  \end{tabular}
  \caption{Properties of the different multifractal time signals and
    associated projected networks.}
  \label{tab:properties}
\end{table}

\begin{figure}[t]
  \includegraphics[width=\linewidth]{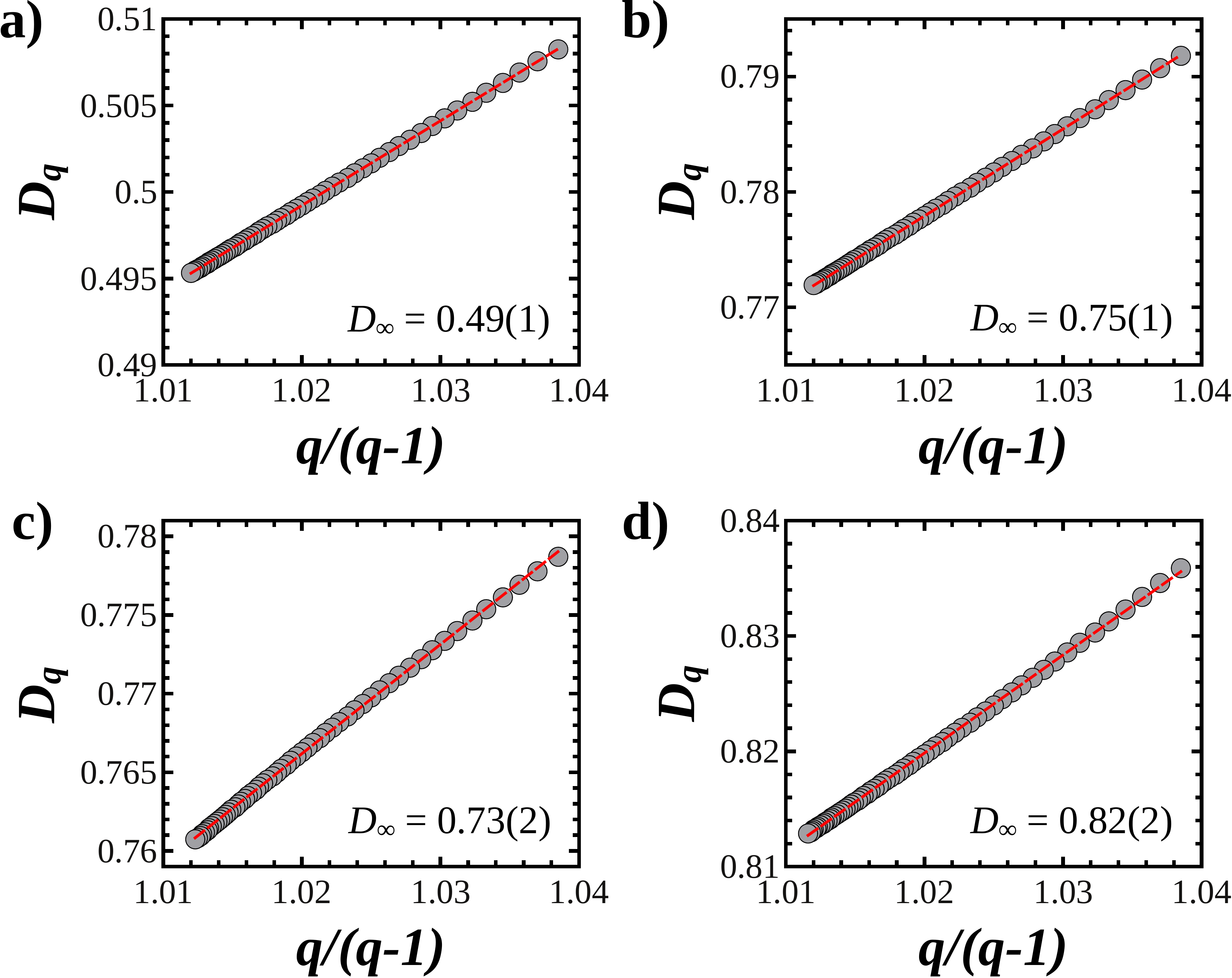}
  \caption{Generalized dimensions $D_q$ as a function of $q/(q-1)$ for
     the logistic map  a),  the Duffing map b), the $d=1$ Henon map
    c) and the $d=2$ Henon map c). The corresponding asymptotic values of
    $D_\infty$ can be extrapolated by means of a linear regression of
    these plots following Eq.~(\ref{eq:21}). The results are listed in
    Table~\ref{tab:properties}. }
  \label{fig:dimensions}
\end{figure}

In Table~\ref{tab:properties} we present a summary of the multifractal
properties of the chaotic time signals considered, computed by using the
box counting formalism described in Sec.~\ref{sec:mult-time-sign}. In
particular, from Eq.~(\ref{eq:9}), we compute the exponent $D_q$ by
performing a linear regression of $\log \sum_i p_i(\eps)$ as a function
of $\log \eps$, for $\eps = M^{1/d}$, in a range of values of $M$
between $10^3$ and $4 \times 10^6$, depending of the particular
attractor. The slope of this regression yields the factor $(q-1)D_q$.
According to Eq.~(\ref{eq:21}), the asymptotic value $D_\infty$ is
obtained by means of linear regressions of $D_q$ as a function of
$q/(q-1)$, performed over suitable intervals of the variable
$q/(q-1)$, see Fig.~\ref{fig:dimensions}.

\begin{figure}[t]
  \includegraphics[width=\linewidth]{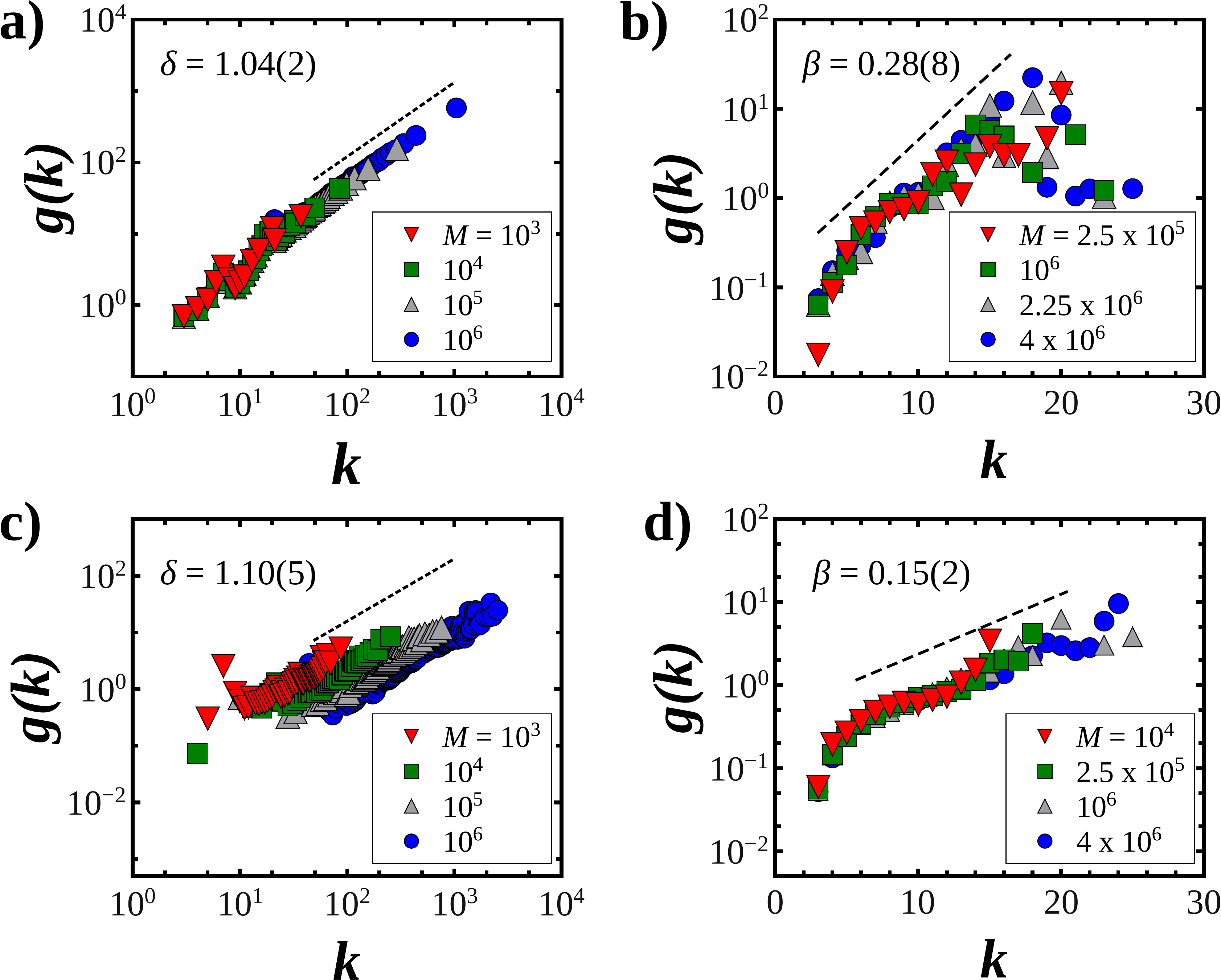}
  \caption{Rescaled average natural measure $\eps^{-D_0} \bar{p}(k)$ as
    a function of the degree $k$ for the logistic map a), the Duffing
    map b), the $d=1$ Henon map c) and the $d=2$ Henon map d). The
    function $\bar{p}(k)$ is computed for different values of $M$,
    considering $n = 1000 \times M$ iterations. Dashed lines represent
    the estimated values of $\delta$ and $\beta$ in each case, see
    Table~\ref{tab:properties}. }
 \label{fig:g_of_k}
\end{figure}

In Fig.~\ref{fig:g_of_k} we plot the natural measure $\bar{p}_\eps(k)$,
averaged over all nodes of degree $k$, as a function of $k$, for
different partitions of the multifractal attractors, i.e. different
$\eps$ (or $M$). In order to check the main assumption in Eq.~(\ref{eq:12}), we
plot the rescaled function $\eps^{-D_0}\bar{p}_\eps(k)$ as a function of the
degree, using the fractal dimensions $D_0$ quoted in
Table~\ref{tab:properties}. In this case, we expect all plots of each map for
different $\eps$ to collapse onto the single universal function
$g(k)$. From Fig.~\ref{fig:g_of_k} we observe, for the logistic, Duffing
and $d=2$ Henon maps, a perfect convergence of the rescaled average natural
measure as a function of $k$, indicating the validity of the assumption
in Eq.~(\ref{eq:12}). In the case of the $d=1$ projection of the Henon
attractor, however, the collapse of $\bar{p}_\eps(k)$ is not
fulfilled. In this particular case, therefore, the predictions made in
Sec.~\ref{sec:topol-prop-proj} are not expected to hold.

From Fig.~\ref{fig:g_of_k}, we also observe that in the $d=1$ cases
(Figs.~\ref{fig:g_of_k}a and \ref{fig:g_of_k}c), both the attractor of
the logistic map and that of the projected Henon map obey an algebraic
behavior, $g(k) \sim k^{\delta}$, while in the $d=2$ cases
(Figs.~\ref{fig:g_of_k}b and \ref{fig:g_of_k}d), both the Duffing and
the Henon systems show an exponential growth, $g(k) \sim e^{\beta k}$.
A linear regression performed on the data in Fig.~\ref{fig:g_of_k}
provides an estimation of the exponents $\delta$ and $\beta$. In
general, it can be noticed how the linear profiles become more defined
and stable while refining the statistics of $g(k)$ by increasing $M$; we
thus fit data obtained with $M=10^{6}$ in the algebraic case and
$M=4 \times 10^6$ for an exponential $g(k)$. We thus find the exponents
$\delta=1.04(2)$ for the logistic map, $\beta=0.28(8)$ for the Duffing
map, while $\delta= 1.10(5)$ and $\beta=0.15(2)$ are obtained for the
$d=1$ and the $d=2$ Henon maps, respectively.

\begin{figure}[t]
  \includegraphics[width=\linewidth]{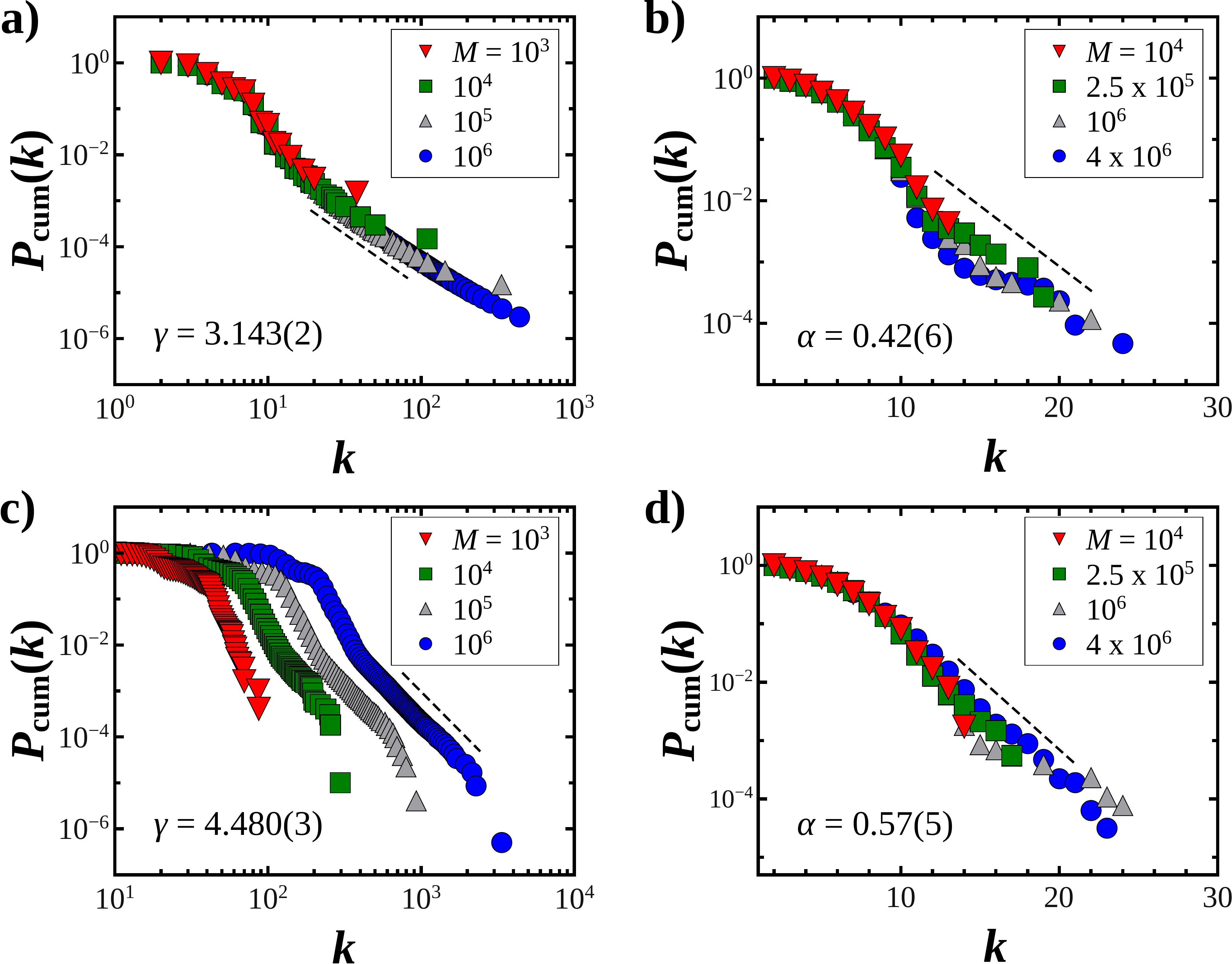}
  \caption{Cumulative degree distributions of the networks projected
    from the logistic map  a),  the Duffing map b), the $d=1$ Henon
    map c) and the $d=2$ Henon map d), computed for different values of $M$ and
    with $n = 1000 \times M$ iterations. Dashed lines represent the
    estimated values of $\gamma$ and/or $\alpha$ of the pertinent case, see
    Table~\ref{tab:properties}.}
  \label{fig:P_of_k}
\end{figure}

In Fig.~\ref{fig:P_of_k} we examine the topology of the projected
networks by plotting the cumulative degree distributions,
$P_\mathrm{cum}(k) = \sum_{q=k}^\infty P(q)$, for different values of
$M$. As predicted in Sec.~\ref{sec:topol-prop-proj}, the networks
characterized by an algebraic growth $g(k)$ (here $d=1$ cases) exhibit
power-law degree distributions. Panels a) and c) show how for different
network sizes (i.e. different values of $M$) all trends converge to a
common power-law distribution characterized by $\gamma = 3.14(1)$ in the
logistic networks and $\gamma = 4.48(1)$ for the $d=1$ Henon
map. Interestingly, in this last case, the prediction of a scale-free
degree distribution holds, despite the fact that Eq.~(\ref{eq:12}) is
not fulfilled. This must attributed to the affect of an algebraic
function $g(k)$, still present in the $d=1$ Henon map.  By contrast, the
networks resulting from the projection of the $d=2$ maps follow
short-tailed degree distributions, compatible with an exponential
behavior $P(k) \sim e^{-\alpha k}$. These plots present a poor
statistics and we can only extrapolate rough values for $\alpha$, namely
$\alpha= 0.42(6)$ and $\alpha=0.57(5)$ for the Duffing and Henon maps,
respectively.

With these values characterizing the multifractal properties of the maps
and the topological properties obtained from the analysis of the
projected transition networks, we can validate our theoretical
framework. We first check the cross-relations given by
Eqs.~(\ref{eq:20}) and (\ref{eq:24}) for exponential and algebraic
cases, respectively. Regarding the algebraic examples we obtain
$(\gamma-1)/\delta = 2.06(7)$ and $D_0/(D_0-D_\infty) = 1.96(5)$ for the
logistic map (see values from Table~\ref{tab:properties}). In this case,
the identity Eq.~(\ref{eq:24}) is well fulfilled within error bars. For
the Henon attractor in $d=1$, we obtain instead
$(\gamma-1)/\delta = 3.2(2)$ and $D_0/(D_0-D_\infty) = 3.7(3)$, again
coinciding within error bars. For exponential cases, the equality
Eq.~(\ref{eq:20}) leads to the comparison $\alpha/\beta = 1.5(6)$ and
$D_0/(D_0-D_\infty) = 2.34(5)$ for the Duffing map, and
$\alpha/\beta = 3.8(8)$ and $D_0/(D_0-D_\infty) =2.9(2)$ for the $d=2$
Henon attractor. In this case, the exponent relations are affected by
stronger errors, but the trend is clearly towards a positive comparison.

\begin{figure}[t]
  \includegraphics[width=\linewidth]{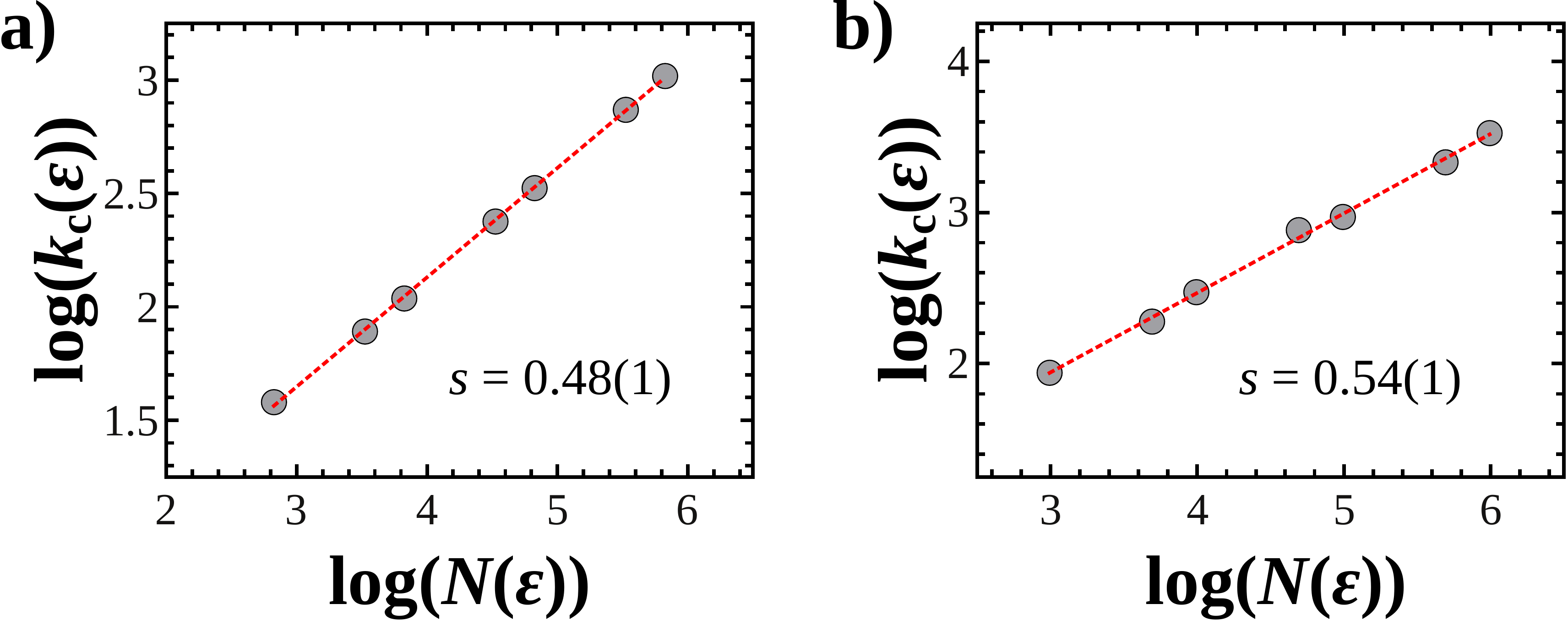}
  \caption{Scaling of the degree cut-off, $k_c(\epsilon)$, estimated as
    the maximum degree in the network, as function of the network size
    $N(\epsilon)$ for the logistic map, a),  and the $d=1$ Henon
    map, b).}
  \label{fig:cut_off}
\end{figure}

\begin{figure}[t]
 \includegraphics[width=0.9\linewidth]{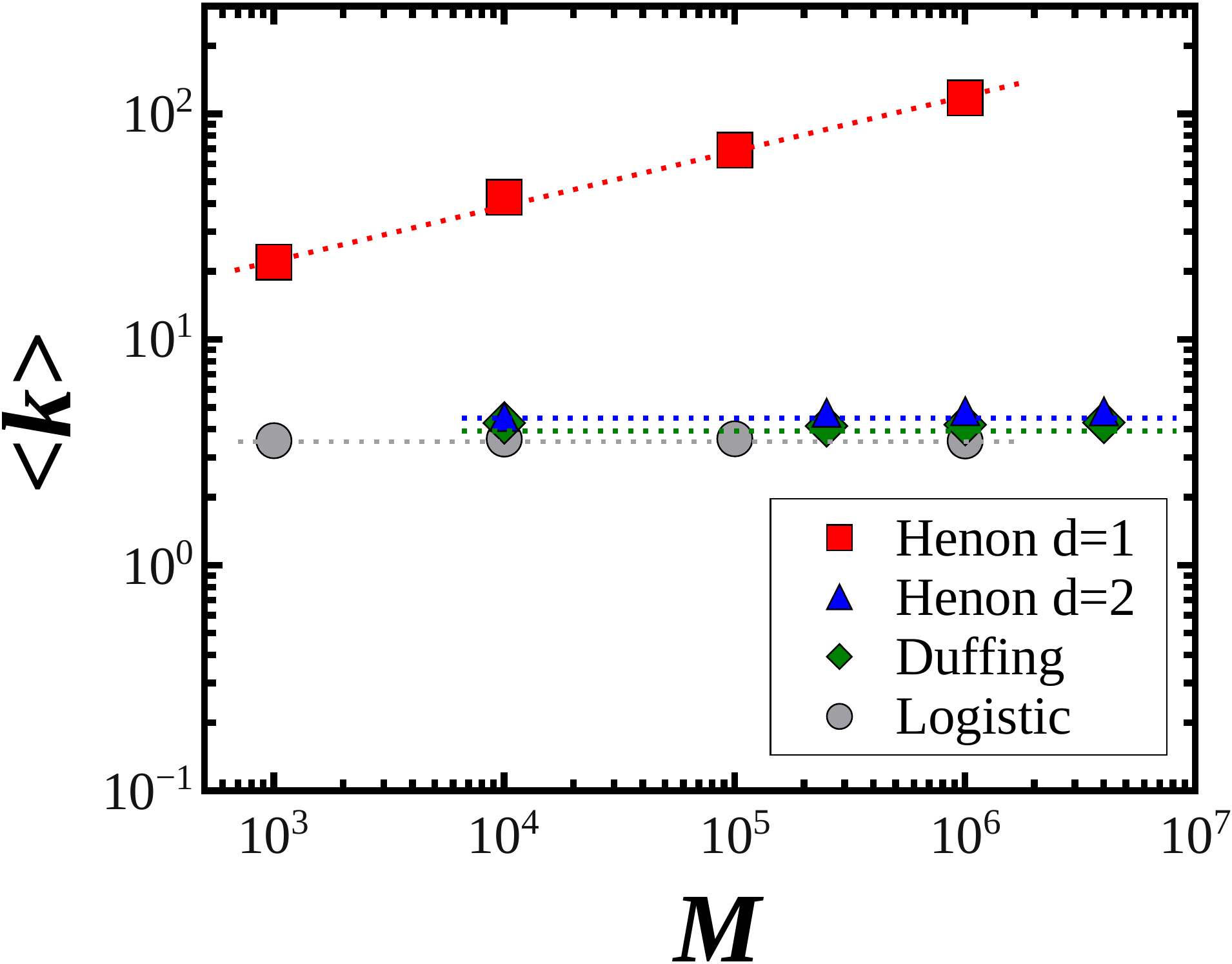}
 \caption{Average degree $\av{k}$ as a function of $M$ for the different
   maps considered. All of them, except the $d=1$ projection of the
   Henon map are essentially independent of the network discretization.}
  \label{fig:average_g}
\end{figure}

Finally, for an algebraic $g(k)$, as observed in the logistic map and
the $d=1$ projection of the Henon attractor, we can proceed to check the
behavior of the maximum degree $k_c(\eps)$ as a function of the network
size $N(\eps)$. We do not consider the relation between degree cut-off
and network size for maps with an exponential $g(k)$, since the very
small span of network sizes obtained does not allow for a determination
of the exponent in relation Eq.~(\ref{eq:3}). Indeed, for the algebraic case,
following Eq.~(\ref{eq:22}), the behavior of the maximum degree is given
by $k_c(\eps) \sim N(\eps)^s$, with an exponent
$s = (1-D_{\infty}/D_{0})/\delta$. The numerical exponents obtained
through a linear regression of $\log k_c(\eps)$ as a function of
$\log N(\eps)$ are $s = 0.48(1)$ and $s = 0.54(1)$ for the logistic and
the $d=1$ Henon map, respectively.  From the values of $\delta$, $D_0$
and $D_\infty$ in Table~\ref{tab:properties}, our theoretical
predictions are $s = 0.48(3)$ and $s = 0.25(3)$ for the logistic and the
$d=1$ Henon map, respectively.  The agreement between numerics and
theory is very good for the logistic map, but completely off in the
$d=1$ Henon case. The disagreement in this last case must be attributed
to the failure of Eq.~(\ref{eq:12}). While the general trend towards a
power-law degree distribution is ensured by the algebraic form of the
function $g(k)$, the lack of the expected scaling with $\eps$ (i.e. the
prefactor) affects the scaling relations deduced from
Eq.~(\ref{eq:12}). In the $d=1$ Henon case yet another of the
approximations made breaks down, namely the assumption of a stable
degree distribution. This fact is checked in Fig.~\ref{fig:average_g},
which shows that the average degree $\av{k}$ of the projected networks
is essentially independent of $\eps$ for all the multifractal time
signals considered, except for the $d=1$ Henon attractor, which exhibits
instead a power-law increasing behavior.
This fact indicates that $d=1$ Henon networks belong to the class of
accelerated networks \cite{Dorogovtsev:2002}.  This increasing average
degree causes the degree distribution to be non stable, introducing an
additional scale, beyond the degree cut-off $k_c$.  Surprisingly,
however, the exponent relation Eq.~(\ref{eq:24}) seems to still be
fulfilled, at least within error bars.

\section{Discussion}
\label{sec:discussion}

In this paper we have investigated the effect of the fractal and
multifractal properties of a temporal signal on the topology of the
corresponding projected network. By combining a transition network
representation with the box counting formalism, we have mapped temporal
signals into networks whose nodes are the boxes partitioning the
attractor of the temporal signal in the phase space, and links are
established between successive pair of boxes between which the signal
jumps. We have developed a mathematical framework connecting network
topology to the multifractal properties of the generating signal. This
formalism allows us to predict the functional form of the network degree
distribution on the basis of the relation, $g(k)$, linking the natural
measure of a box with the associated node degree.  We have focused on
the prototypical and general cases of an exponential and an algebraic
growth $g(k)$, showing that the latter results in power law degree
distributions whose exponent $\gamma$ is controlled by the multifractal
exponents of the generating signal. We have verified the validity of our
approach through extensive numerical simulations, highlighting the
excellent agreement observed in many cases, and discussing in detail the
reasons why in some cases (e.g., the Henon map in $d=1$) the numerical
experiments depart from theoretical predictions.

In particular, we could conclude that a sufficient condition to obtain a
scale-free topology is that the natural measure of a box must increase
with the degree of the associated node in a algebraic fashion. In our
numerical experiments we have observed that this condition is fulfilled
in multifractal attractors in $d=1$ with fractal dimension $D_0=1$. This
fact leads us to conjecture that scale-free networks can be observed in
general multifractal time series in which the fractal dimension is equal
to the euclidean dimension of the embedding phase space.

Our work extends existing approaches bridging time series analysis and
network science by addressing the ubiquitous case of signals exhibiting
multifractal properties. By doing this, it enriches the set of
interpretative tools available for a better characterization of
empirical time-series. For this reason, we can envisage that it will be
of interest also to the growing community of interdisciplinary
researchers studying natural time series through the lenses of network
science.

\begin{acknowledgments}
  M.A.B. acknowledges financial support from
  FRS-FNRS. R.P.-S. acknowledges financial support from the Spanish
  MINECO, under projects FIS2013-47282-C2-2 and FIS2016-76830-C2-1-P,
  and EC FET-Proactive Project MULTIPLEX (Grant No. 317532).
  R.P.-S. acknowledges additional financial support from ICREA Academia,
  funded by the Generalitat de Catalunya.
\end{acknowledgments}

\end{document}